\documentclass[prl,twocolumn,superscriptaddress,showpacs]{revtex4}

\usepackage{amsmath,amssymb,ulem,color}
\usepackage{graphicx}

\renewcommand{\l}{\left}
\renewcommand{\r}{\right}
\renewcommand{\b}{\beta}
\newcommand{\al}{\alpha}
\newcommand{\q}{\quad}
\newcommand{\ep}{\varepsilon}
\newcommand{\Ref}[1]{(\ref{#1})}
\newcommand{\lam}[1]{\lambda^{{#1}}_\ep(x)}

\newcommand{\Z}{{\mathbb{Z}}}

\begin{document}

\title{Non-meanfield deterministic limits in chemical reaction
  kinetics far from equilibrium}

\author{R. E. Lee DeVille} 

\affiliation{Courant Institute of Mathematical Sciences, New York
University, New York, NY 10012}

\author{Cyrill B. Muratov}

\affiliation{Department of Mathematical Sciences, New Jersey Institute
of Technology, Newark, NJ 07102}

\author{Eric Vanden-Eijnden} 

\affiliation{Courant Institute of Mathematical Sciences, New York
University, New York, NY 10012}

\date{\today}

\begin{abstract}
  A general mechanism is proposed by which small intrinsic
  fluctuations in a system far from equilibrium can result in nearly
  deterministic dynamical behaviors which are markedly distinct from
  those realized in the meanfield limit. The mechanism is demonstrated
  for the kinetic Monte-Carlo version of the Schnakenberg reaction
  where we identified a scaling limit in which the global
  deterministic bifurcation picture is fundamentally altered by
  fluctuations.  Numerical simulations of the model are found to be in
  quantitative agreement with theoretical predictions.
\end{abstract}

\pacs{05.40.-a, 82.40.Bj, 82.39.-k, 02.50.Fz}

\maketitle


On a mesoscopic level, large-scale dynamical systems are always
subject to fluctuations, either due to the perturbations coming from
the random environment (extrinsic noise), or due to the fundamental
randomness of the underlying physical processes (intrinsic noise)
\cite{fizkiny,gardiner}. 
Often, these fluctuations cannot be neglected.  For instance, small
noise can alter the system behavior in essential ways by producing
activated processes such as rare barrier crossing events
\cite{freidlin,gardiner,hanggi90}.

Perhaps even more surprisingly, rare events coupled to slow
deterministic dynamics may result in behaviors that are dramatically
different from the dynamics in the absence of noise and yet remain
essentially deterministic \cite{freidlin01sd}. By now a classical
example of this phenomenon is \emph{stochastic resonance}, whereby
under appropriate conditions the noise can induce an essentially
all-or-none response to a driving frequency \cite{wellens04}.  Another
example is the recently discovered phenomenon of self-induced
stochastic resonance (SISR) in which the external noise plays a
constructive role in producing new non-meanfield behaviors
\cite{freidlin01,mve:pd05,dvm05}. In SISR the interplay between the
slow Arrhenius time scale \cite{gardiner,freidlin,hanggi90} of rare
events and the slow deterministic time scale is achieved by tuning
appropriately the amplitude of the external noise (making both these
time scales long ensures the deterministic character of the observed
behavior).

On the other hand, one might ask whether the same type of interplay is
possible in systems in which the fluctuations are generated
intrinsically. One important class of such systems are kinetic Monte
Carlo (KMC) schemes \cite{gillespie76,gillespie77} for chemical and
biochemical reactions
\cite{schnakenberg76,white00,vilar02,paulsson04}, in which randomness
is due to the underlying stochastic jump processes \cite{gardiner}.
In KMC, the level of noise is not a free parameter, but is determined
by all the processes combined. Therefore, it is not clear \textit{a
  priori} that the right balance can in fact be achieved.

In this Letter, we show that the non-trivial interplay between rare
events and slow dynamics is indeed possible in systems far from
equilibrium with intrinsic randomness such as KMC. We do so by
identifying distinguished limits in which the dynamics of the system
becomes completely deterministic and at the same time remains distinct
from what is obtained in the meanfield limit. The mechanism is first
explained by means of a general model, and then demonstrated to be
feasible in a specific example of an autocatalytic reaction KMC
scheme.

Our general model is a ``drift-or-jump'' dynamical system whose state
at any moment is described by a vector $y \in \mathbb R^m$. During an
infinitesimal time interval $dt$ the system will move by $dy =
g_\al(y) dt$ to a new location $y + dy$, or jump instantaneously to a
new position $p(y) \in \mathbb R^m$ with probability $k_\ep(y) dt$.
Here $g_\al(y)$ is a given vector field, $k_\ep(y)$ is a jump rate,
and $p(y)$ is a mapping which, for simplicity, we will assume to be
one-to-one. The Chapman-Kolmogorov equation for the probability
density $\rho(y,t)$ of this Markov process is \cite{gardiner}
\begin{equation}
  \label{eq:rho}
  \begin{aligned}
    {\partial \over \partial t} \rho(y, t) & = - {\partial \over
      \partial y} ( g_\al(y) \rho(y, t) )\\ & \quad - k_\ep(y) \rho(y, t)
    + k_\ep(p^{-1}(y)) \rho(p^{-1}(y), t).
  \end{aligned}
\end{equation}
In the following, we are interested in a particular situation in which
the drift is ``slow'' and the jumps are ``rare'', quantified by two
small dimensionless parameters, $\al$ and $\ep$, respectively. More
precisely, we assume that 
\begin{eqnarray}
  \label{eq:k}
  g_\al(y) = \al g(y), \qquad k_\ep(y) = \nu(y) \exp\{ -\ep^{-1} V(y)\},
\end{eqnarray}
i.e. $\al$ characterizes the slow time scale of the drift generated by
$g_\al$ relative to some reference $O(1)$ time scale, and $\ep$ is the
intensity of the jumps. Importantly, we assumed that the rate
$k_\ep(y) $ is in Arrhenius form, with $V(y)$ playing the role of a
``barrier height'' to be crossed in the event of a jump and $\nu(y)$
being a dimensional rate prefactor assumed to be $O(1)$ on the
reference time scale.


If we let $\ep\to0$ in Eq.~(\ref{eq:rho}) with all other parameters
fixed, then the last two terms in Eq.~(\ref{eq:rho}) disappear, and
the limiting dynamics reduces to the deterministic motion $\dot y =
\alpha g(y)$, the meanfield limit.
However, as we show now, more interesting behaviors occur if we let
$\ep \to 0$ and $\al\to0$ simultaneously on some specific sequence and
make some extra assumptions. For instance, suppose that (i) the map
$p(y)$ satisfies $\forall y\in \mathbb R^m$: $V(p(y))> V(y)$ (i.e.
$y$ always jumps to a state where the jump rate is smaller), and (ii)
$\forall y\in \mathbb R^m$: $\nabla V(y) \cdot g(y)<0$ (i.e. the
deterministic drift drives the system toward regions of higher jump
rate). Rescale time and introduce $\beta$ as
\begin{eqnarray}
  \label{eq:beta}
  \tau = \al t, \qquad \beta = \ep \ln \al^{-1},
\end{eqnarray}
and observe that in Eq.~(\ref{eq:rho}) written in the new time scale
the jump rate is $k(y) \equiv \al^{-1} k_\ep(y) = \nu(y) \exp\{
\ep^{-1} (\beta - V(y) \}$. Therefore, if we let $\alpha,\ep\to0$ in a
way that $\beta = O(1)$ is kept fixed, then
\begin{equation}
  \label{eq:limk}
  k(y) \to  
  \begin{cases}
    0, & y \in \Omega_\beta, \\
    \infty, & y \not\in \bar \Omega_\beta,
  \end{cases}
\end{equation}
where $\Omega_\beta \subset \mathbb R^m$ is the region where $V(y) >
\beta$.  So the following will happen: Supposing that the system
starts from $y \in \Omega_\beta$, it will drift deterministically
toward the boundary of this region, $\partial \Omega_\beta$, where
$V(y)=\beta$. Once it reaches $\partial \Omega_\beta$, an
instantaneous jump will occur to a new location $p(y) \in
\Omega_\beta$ by assumption above, and the process will repeat itself
indefinitely. The resulting dynamics in this limit is deterministic
(since both the drift and the jump outcome are prescribed), but it is
{\em strongly} non-meanfield, since it is very different from the
solution of $d y/d \tau = g(y)$
(see also \cite{mve:pd05,dvm05,freidlin01}). We also note that for
small but finite $\al, \ep$ this dynamics should be augmented by a
boundary layer analysis of Eq.~(\ref{eq:rho}) near $\partial
\Omega_\beta$.

In the model above we postulated Eq.~(\ref{eq:rho}) and made some
specific assumptions about the terms in this equation. Next we show
that this equation can in fact be \textit{derived} in a particular
example in which it is not \textit{a priori} obvious. We take the KMC
scheme for the Schnakenberg reaction~\cite{schnakenberg79}:
\begin{equation}
  \begin{split}\label{eq:S}
    S_1 &\xrightarrow{k_1} X,\\
    X &\xrightarrow{k_2} \mbox{Products},\\
    2X + Y &\xrightarrow{k_3} 3X,\\
    S_2 &\xrightarrow{k_4} Y,
  \end{split}
\end{equation}
which is a classical autocatalytic reaction scheme exhibiting limit
cycle oscillations and capturing a number of essential non-equilibrium
features of more realistic chemical and biochemical reactions (see,
e.g., \cite{goldbeter}).

In the KMC version of the Schnakenberg reaction the state of the
system at any time is given by the pair $(X,Y)$ of integers
corresponding to the numbers of molecules of the respective species.
To identify the fast/slow dynamics, we first introduce the rescaling
(we absorb $S_1$ and $S_2$ into the rate constants)
\begin{eqnarray}
  \label{eq:scale}
  x = k_1^{-1} X , \qquad y = (k_3 k_1^2 / k_4) \, Y
\end{eqnarray}
and the dimensionless quantities
\begin{eqnarray}
  \label{eq:dim}
  \al = k_1^2 k_3 / k_2^3, \quad \ep = k_2 / k_1, \quad A = k_4 /
  k_1. 
\end{eqnarray}
After rescaling, the Chapman-Kolmogorov equation for the probability
density function $\rho(x,y,t)$, where $(x,y) \in \ep \Z_+ \times
(\ep\al/A) \Z_+$, of this Markov process is
\begin{eqnarray}
  \label{eq:KMCscaled}
  {\partial \rho(x,y,t)\over \partial t} = 
  \ep^{-1} [\rho(x-\ep,y,t)-\rho(x,y,t)\nonumber \\
  + (x+\ep)\rho(x+\ep,y,t) - x \rho(x,y,t) \nonumber \\
  + A (x - \ep)(x-2\ep) (y+\ep\al/A)
  \rho(x-\ep,y+\ep\al/A,t) \\
  - Ax(x-\ep)y\rho(x,y,t) \nonumber \\
  + A \rho(x,y-\ep\al/A,t) - A \rho(x,y,t)]. \nonumber 
\end{eqnarray}
%
%
Taking the limit $\ep \to 0$ in Eq.~(\ref{eq:KMCscaled}) with all
other parameters fixed, we obtain the deterministic process described
by the mass action law (the meanfield limit):
\begin{equation}
  \label{eq:ODEscaled}
  \begin{cases}
    \dot x = 1 - x + A x^2y,\\
    \dot y = \al (1 - x^2y),
  \end{cases}
\end{equation}
where now we measure time in the units of $k_2^{-1}$. From this one
can see that the constant $\al$ measures the time scale separation
ratio between $x$ and $y$, and the constant $A$ controls the location
of the unique fixed point
\begin{eqnarray}
  \label{eq:xy0}
  x_0 = 1 + A, \qquad y_0 = (1 + A)^{-2}.
\end{eqnarray}

\begin{figure}[t]
  \centerline{\includegraphics[width=3.25in]{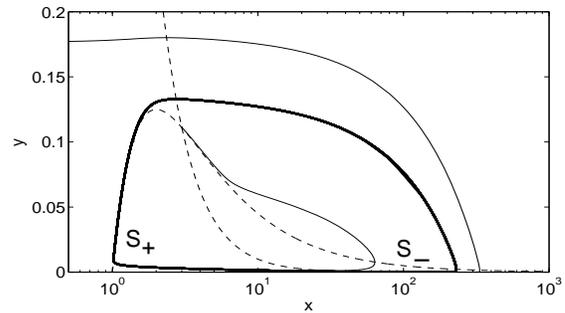}}
  \caption{Phase plane trajectories for Eqs.~(\ref{eq:ODEscaled}) with
    $A = 2$ and $\al = 10^{-3}$. The limit cycle is indicated with a
    solid loop.}
  \label{fig:ODE}
\end{figure}

For $\al$ small enough Eqs.~(\ref{eq:ODEscaled}) exhibit a limit cycle
when $A > 1$, i.e.  when the fixed point $(x_0, y_0)$ lies on the
unstable branch $S_+$, where $x = x_+(y)$,
\begin{eqnarray}
  \label{eq:nul}
  x_\pm(y) = (1 \pm \sqrt{1 - 4 A y})/(2 A y),
\end{eqnarray}
of the $x$-nullcline (Fig. \ref{fig:ODE}). In contrast, there is no
limit cycle when $A < 1$. In this case the fixed point lies on the
stable branch $S_-$, where $x = x_-(y)$, of the nullcline and this
fixed point is stable and globally attracting. During the slow motion,
$x$ remains close to $x_-(y)$, and so to leading order in
$\alpha\ll1$, $y$ satisfies $\dot y = \al g(y)$ with
\begin{eqnarray*}
g(y) = 1 - x_-^2(y) y.
\end{eqnarray*}

\begin{figure*}[t]
  \centerline{\includegraphics[width=5.5in]{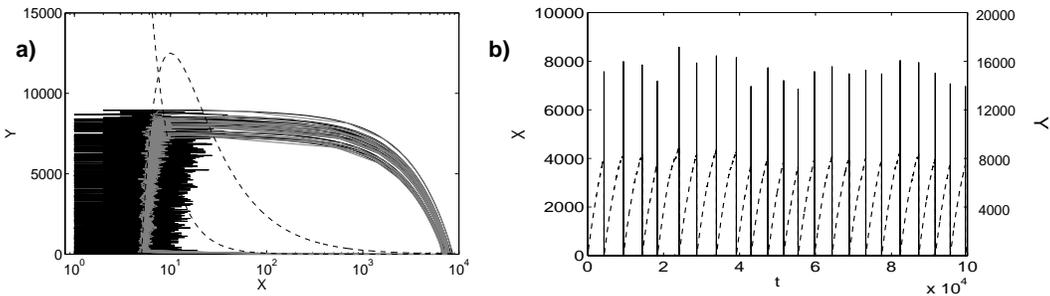}}
  \caption{Results of the KMC simulation of Eq.~\Ref{eq:S}, with $k_1
    = 5, k_2 = 1, k_3 = 4\times 10^{-6}, k_4 = 2.5$.  In (a), we plot
    a sample trajectory in $(X,Y)$-space in black.  Over this we have
    plotted in gray the running average over 1000 steps of the
    simulation, to show the typical size of the fluctuation (the
    fluctuations in the raw trajectory are exaggerated by rare large
    deviations).  In (b), we plot $X$ and $Y$ versus time.  
}
  \label{fig:trace}
\end{figure*}

Now we turn to the analysis of the model for $\al, \ep \ll 1$ with
$\beta = O(1)$, following the general discussion earlier and show that
a non-meanfield deterministic behavior emerges in this case. To obtain
the transition rate $k_\ep(y)$, we need to consider the escapes from the
vicinity of $S_-$ with $y$ frozen. 
Setting $\al=0$ in Eq.~(\ref{eq:KMCscaled}), one sees that this
amounts to studying a one-dimensional jump process in~$x$ with right
and left jump rates:
\begin{equation}
  \label{eq:1drates}
  \lam+ = 1 + A x (x - \ep) y, \q \lam- = x.
\end{equation}
For $\ep \ll 1$ the trajectory needs to reach the vicinity of $S_+$ in
order to escape the basin of attraction of $S_-$. The corresponding
rate of this escape event gives the $k_\ep(y)$ to use in the
equivalent of Eq.~(\ref{eq:rho}), and it can be determined by
generalizing the classical analysis of Kramers \cite{gardiner} for the
one-dimensional jump process with rates from Eq.~(\ref{eq:1drates})
\cite{dmv}.
This gives an expression for $k_\ep(y)$ which is in the form of
Eq.~(\ref{eq:k}) where (see also \cite{shwartz})
\begin{eqnarray}
  \label{eq:bar}
  V(y) = \int_{x_-(y)}^{x_+(y)}
  \ln\l(\frac{\lambda^-_0(x)}{\lambda^+_0(x)}\r)\,dx 
\end{eqnarray}
and
\begin{eqnarray}
  \label{eq:nu}
  \nu(y) =  {2 A y \sqrt{ 1 - 4 A y} \over \pi (1 + \sqrt{1 - 4 A y}
    \, )^2
  }.  
\end{eqnarray}

The function $V(y)$ in Eq.~(\ref{eq:bar}) can be easily computed in
closed form. In particular, $V(y)$ is a monotonically decreasing
function of $y$ for fixed $A$.  It follows that, given any $\beta>
\beta_c(A) = V(y_0)$, we have $\beta = V(y_\star)$ for some $y_\star =
y_\star(\beta, A)$ with $0 < y_\star < y_0$. As a result, in the limit
$\al, \ep \to 0$ with $\beta > \beta_c(A)$ fixed, the trajectory will
jump precisely at $y = y_\star$ consistent with Eq.~(\ref{eq:limk})
(here $\Omega_\beta = \{y< y_\star\}$ and $\partial \Omega_\beta =
\{y_\star\}$).

After escape the process undergoes an excursion governed by
Eq.~(\ref{eq:ODEscaled}) (instantaneous on the current time-scale),
similar to the one considered in Ref.  \cite{mve:pd05}, after which it
returns to $S_-$ at $p(y)=0$. Thus, the non-meanfield behavior
predicted from Eq.~(\ref{eq:rho}) precisely arises in the present
example and is an instance of SISR due to intrinsic noise. The
observed non-meanfield behavior is a limit cycle with period (in units
of $\al^{-1}$)
\begin{equation}\label{eq:defofT}
  T(A,\b) = \int_{0}^{y_\star} \frac{dy}{g(y)},
\end{equation}
which is controlled by $\beta$.

For small but finite $\al$ and $\ep$ the escape region will be
smeared. To compute the deviations from the deterministic limit, we
expand $V(y)$ in Taylor series in $y - y_\star$. Introduce $z =
\ep^{-1} (y - y_\star - \mathit{\Delta} y_\star)$, where
$\mathit{\Delta} y_\star$ is a deterministic correction to $y_\star$
to be determined. Then, after a straightforward computation we obtain
that to leading order $ \mathit{\Delta} y_\star = \ep \ln \ep^{-1} / |
V'(y_\star)|$, where $V' = dV/dy$, and the boundary layer solution
$\rho_0(z)$ valid for $|z| \ll \ep^{-1}$ is
\begin{eqnarray}
  \rho_0(z) = C \exp \left\{ - {\nu(y_\star) \over g(y_\star)
      | V'(y_\star)|} e^{| V'(y_\star)|
      z} \right\}.  
\end{eqnarray}
With $C = 1$ this is also the probability that the trajectory has not
jumped yet at $y = y_\star + \mathit{\Delta} y_\star + \ep z$. Using
this, we can compute the average jump-off point $\langle y
\rangle$. Evaluating the necessary integrals, we finally obtain (to
leading order in $\ep$)
\begin{eqnarray}
  \label{eq:yav}
  \langle y \rangle - y_\star = {\ep \over |
    V'(y_\star)|} \ln   \left( { g(y_\star) |
      V'(y_\star)| \over \ep \nu(y_\star) 
      e^\gamma} \right),
\end{eqnarray}
where $\gamma \simeq 0.5772$ is the Euler constant. Similarly, the
standard deviation of the jump-off point is (to leading order)
\begin{eqnarray}
  \label{eq:dev}
  \sqrt{ \langle y^2 \rangle - \langle y \rangle^2 } = {\pi \ep
    \over | V'(y_\star)| \sqrt{6} }.
\end{eqnarray}
Note that the deterministic correction $\mathit{\Delta} y_\star =
\langle y \rangle - y_\star $ contains a large logarithm and therefore
always dominates the fluctuating contribution. Also, since the balance
needed to obtain $\beta = O(1)$ implies that $\ep = O(\ln \al^{-1})$,
this, in turn, implies that $\mathit{\Delta} y_\star = O(\ln \ln
\al^{-1})$ and, therefore, gives a noticeable correction to $y_\star$
unless $\al$ is unrealistically small. We also point out that this
results in the noticeable shift $\mathit{\Delta} T = \mathit{\Delta}
y_\star / g(y_\star)$ of the period of the limit cycle.

\begin{figure}
  \centerline{\includegraphics[width=3.25in]{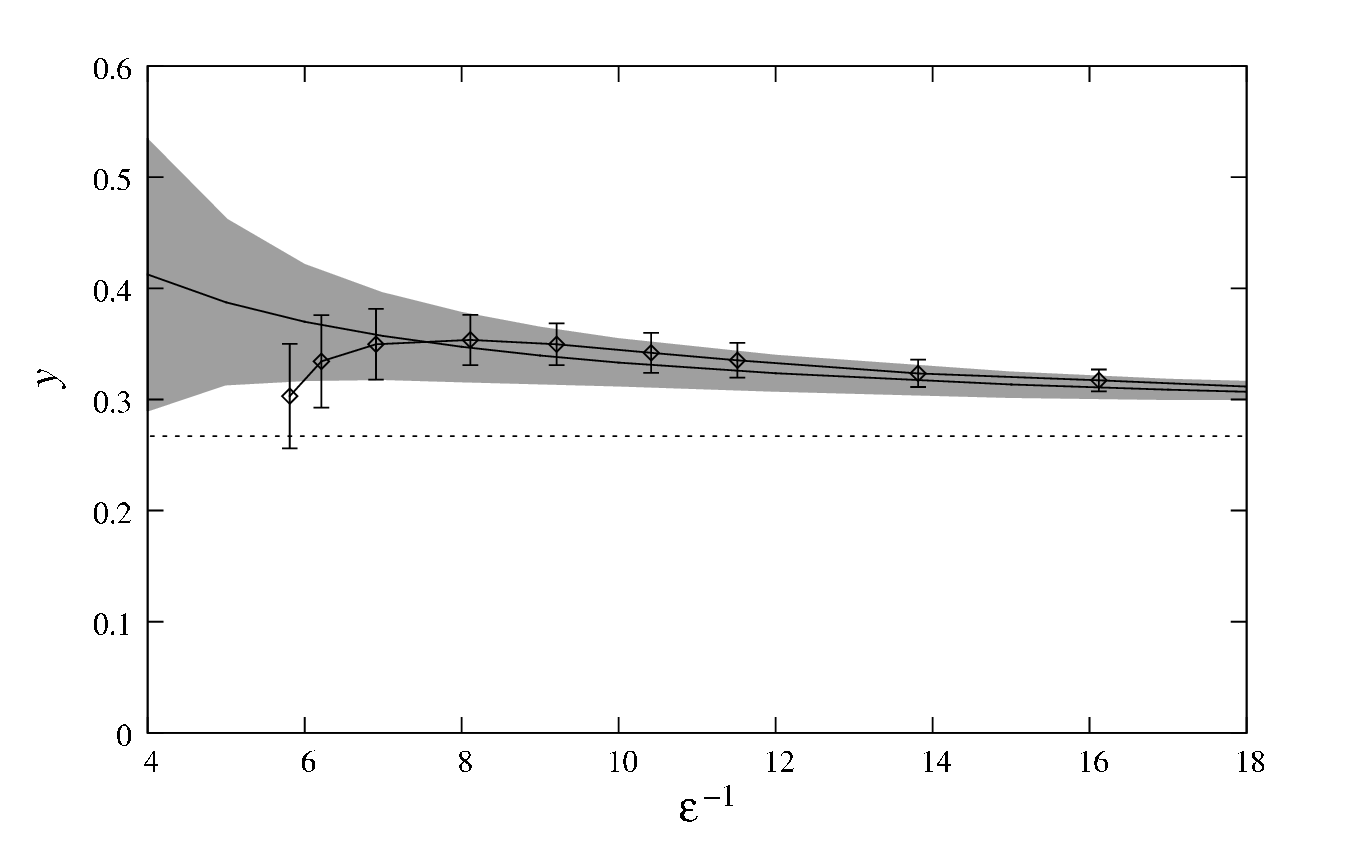}}
  \caption{Mean and standard deviation of the jump-off point $y$ from
    the KMC simulations (data points) and theory (gray).  
    The dotted line shows the value of $y_\star$.  }
  \label{fig:periods}
\end{figure}

\begin{figure}[b]
  \centerline{\includegraphics[width=3.2in]{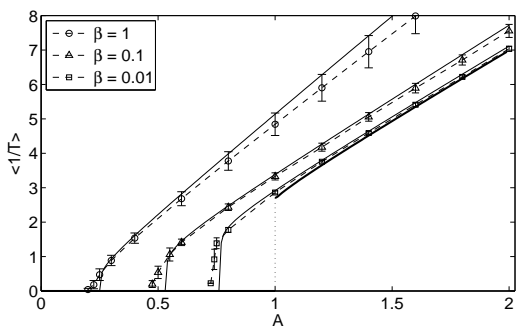}}
  \caption{Bifurcation diagram for the noise-induced limit
    cycle. Points are results of the simulations with $\al = 10^{-5}$,
    thin lines are theoretical predictions for the corresponding
    values of $\beta$. 
  }
  \label{fig:MSD}
\end{figure}

We now check the validity of the picture presented above with KMC
simulations of the Schnakenberg model. Figure \ref{fig:periods} shows
the results of a simulation for $\al = 10^{-4}$, $A = 0.5$, and $\ep =
0.2$, a parameter set at which Eq.~(\ref{eq:ODEscaled}) has no limit
cycle. This figure clearly shows a noise-induced coherent limit cycle
due to SISR \cite{mve:pd05}. Furthermore, by choosing parameters
appropriately, we can make this motion more and more coherent: in
Fig. \ref{fig:periods} we plot the mean and standard deviation (as
errorbars) of the jump-off point $y$ (see Fig. \ref{fig:trace}a).
Here we take the parameters corresponding to $\b = 1$, $A = 0.5$ and
vary $\ep$.  Constraining the rate constants in this way and letting
$\ep$ decrease, we can improve coherence of the SISR limit cycle. Note
that the average jump-off point for the noise-induced limit cycle
differs rather significantly from $y_\star = 0.267$ predicted by the
asymptotic theory due to the very slow convergence of the
latter. However, including the fluctuation corrections from
Eq.~(\ref{eq:yav}) results in excellent quantitative agreement between
the simulation data and theory.

The effect of the intrinsic noise is demonstrated most dramatically by
looking at the global bifurcation picture for the noise-induced limit
cycle. In the absence of the noise the limit cycle appears via a Hopf
bifurcation at $A = 1 + O(\al)$, and in the limit $\al \to 0$ appears
in a discontinuous fashion at $A = 1$ (solid line in
Fig. \ref{fig:MSD}). The noise changes this qualitatively: in
Fig. \ref{fig:MSD} we plotted the average frequency of the limit cycle
from the simulations for several values of $\beta$ with $\al =
10^{-5}$ fixed. The period of the limit cycle shows $O(1)$ deviation
from the meanfield behavior (solid line) for {\em all} values of
$A$. The nature of the bifurcation is also altered: it is now creating
an infinite period orbit. This is consistent with our theoretical
prediction that (asymptotically) the limit cycle is born at $A = A_c <
1$, where $A_c$ solves $\beta = \beta_c(A)$. The predicted limit cycle
period also shows excellent agreement with the numerics in
Fig. \ref{fig:MSD} (thin lines).

In conclusion, we have demonstrated that intrinsic fluctuations in
systems far from equilibrium possessing fast/slow dynamics can have a
profound effect on the observed dynamics, producing strongly
non-meanfield, yet essentially deterministic dynamical behaviors.

The authors acknowledge partial support by NIH R01 GM076690 (C.B.M),
NSF DMS02-09959, NSF DMS02-39625, and ONR N00014-04-1-0565 (E.V.-E.)
grants. This work was performed while E.V.-E. was visiting UC Berkeley
on a Visiting Miller Research Professor.

\bibliography{../mura,../stat,../nonlin,../bio}

\begin{thebibliography}{19}
\expandafter\ifx\csname natexlab\endcsname\relax\def\natexlab#1{#1}\fi
\expandafter\ifx\csname bibnamefont\endcsname\relax
  \def\bibnamefont#1{#1}\fi
\expandafter\ifx\csname bibfnamefont\endcsname\relax
  \def\bibfnamefont#1{#1}\fi
\expandafter\ifx\csname citenamefont\endcsname\relax
  \def\citenamefont#1{#1}\fi
\expandafter\ifx\csname url\endcsname\relax
  \def\url#1{\texttt{#1}}\fi
\expandafter\ifx\csname urlprefix\endcsname\relax\def\urlprefix{URL }\fi
\providecommand{\bibinfo}[2]{#2}
\providecommand{\eprint}[2][]{\url{#2}}

\bibitem[{\citenamefont{Lifshits and Pitaevskii}(1981)}]{fizkiny}
\bibinfo{author}{\bibfnamefont{E.~M.} \bibnamefont{Lifshits}} \bibnamefont{and}
  \bibinfo{author}{\bibfnamefont{L.~P.} \bibnamefont{Pitaevskii}},
  \emph{\bibinfo{title}{Physical kinetics}} (\bibinfo{publisher}{Pergamon
  Press}, \bibinfo{address}{Oxford}, \bibinfo{year}{1981}).

\bibitem[{\citenamefont{Gardiner}(1985)}]{gardiner}
\bibinfo{author}{\bibfnamefont{C.~W.} \bibnamefont{Gardiner}},
  \emph{\bibinfo{title}{Handbook of stochastic methods.}}
  (\bibinfo{publisher}{Springer-Verlag}, \bibinfo{address}{Berlin},
  \bibinfo{year}{1985}).

\bibitem[{\citenamefont{Freidlin and Wentzell}(1984)}]{freidlin}
\bibinfo{author}{\bibfnamefont{M.~I.} \bibnamefont{Freidlin}} \bibnamefont{and}
  \bibinfo{author}{\bibfnamefont{A.~D.} \bibnamefont{Wentzell}},
  \emph{\bibinfo{title}{Random Perturbations of Dynamical Systems}}
  (\bibinfo{publisher}{Springer}, \bibinfo{address}{New York},
  \bibinfo{year}{1984}).

\bibitem[{\citenamefont{H\"anggi et~al.}(1990)\citenamefont{H\"anggi, Talkner,
  and Borkovec}}]{hanggi90}
\bibinfo{author}{\bibfnamefont{P.}~\bibnamefont{H\"anggi}},
  \bibinfo{author}{\bibfnamefont{P.}~\bibnamefont{Talkner}}, \bibnamefont{and}
  \bibinfo{author}{\bibfnamefont{M.}~\bibnamefont{Borkovec}},
  \bibinfo{journal}{Rev. Mod. Phys.} \textbf{\bibinfo{volume}{62}},
  \bibinfo{pages}{251} (\bibinfo{year}{1990}).

\bibitem[{\citenamefont{Freidlin}(2001{\natexlab{a}})}]{freidlin01sd}
\bibinfo{author}{\bibfnamefont{M.~I.} \bibnamefont{Freidlin}},
  \bibinfo{journal}{Stoch. Dyn.} \textbf{\bibinfo{volume}{1}},
  \bibinfo{pages}{261} (\bibinfo{year}{2001}{\natexlab{a}}).

\bibitem[{\citenamefont{Wellens et~al.}(2004)\citenamefont{Wellens, Shatokhin,
  and Buchleitner}}]{wellens04}
\bibinfo{author}{\bibfnamefont{T.}~\bibnamefont{Wellens}},
  \bibinfo{author}{\bibfnamefont{V.}~\bibnamefont{Shatokhin}},
  \bibnamefont{and}
  \bibinfo{author}{\bibfnamefont{A.}~\bibnamefont{Buchleitner}},
  \bibinfo{journal}{Rep. Prog. Phys.} \textbf{\bibinfo{volume}{67}},
  \bibinfo{pages}{45} (\bibinfo{year}{2004}).

\bibitem[{\citenamefont{Freidlin}(2001{\natexlab{b}})}]{freidlin01}
\bibinfo{author}{\bibfnamefont{M.~I.} \bibnamefont{Freidlin}},
  \bibinfo{journal}{J. Stat. Phys.} \textbf{\bibinfo{volume}{103}},
  \bibinfo{pages}{283} (\bibinfo{year}{2001}{\natexlab{b}}).

\bibitem[{\citenamefont{Muratov et~al.}(2005)\citenamefont{Muratov,
  Vanden~Eijnden, and E}}]{mve:pd05}
\bibinfo{author}{\bibfnamefont{C.~B.} \bibnamefont{Muratov}},
  \bibinfo{author}{\bibfnamefont{E.}~\bibnamefont{Vanden~Eijnden}},
  \bibnamefont{and} \bibinfo{author}{\bibfnamefont{W.}~\bibnamefont{E}},
  \bibinfo{journal}{Physica D} \textbf{\bibinfo{volume}{210}},
  \bibinfo{pages}{227} (\bibinfo{year}{2005}).

\bibitem[{\citenamefont{DeVille et~al.}(2005)\citenamefont{DeVille,
  Vanden~Eijnden, and Muratov}}]{dvm05}
\bibinfo{author}{\bibfnamefont{R.~E.~L.} \bibnamefont{DeVille}},
  \bibinfo{author}{\bibfnamefont{E.}~\bibnamefont{Vanden~Eijnden}},
  \bibnamefont{and} \bibinfo{author}{\bibfnamefont{C.~B.}
  \bibnamefont{Muratov}}, \bibinfo{journal}{Phys. Rev. E}
  \textbf{\bibinfo{volume}{72}}, \bibinfo{pages}{031105}
  (\bibinfo{year}{2005}).

\bibitem[{\citenamefont{Gillespie}(1976)}]{gillespie76}
\bibinfo{author}{\bibfnamefont{D.~T.} \bibnamefont{Gillespie}},
  \bibinfo{journal}{J. Comput. Phys.} \textbf{\bibinfo{volume}{22}},
  \bibinfo{pages}{403} (\bibinfo{year}{1976}).

\bibitem[{\citenamefont{Gillespie}(1977)}]{gillespie77}
\bibinfo{author}{\bibfnamefont{D.~T.} \bibnamefont{Gillespie}},
  \bibinfo{journal}{J. Stat. Phys.} \textbf{\bibinfo{volume}{16}},
  \bibinfo{pages}{311} (\bibinfo{year}{1977}).

\bibitem[{\citenamefont{Schnakenberg}(1976)}]{schnakenberg76}
\bibinfo{author}{\bibfnamefont{J.}~\bibnamefont{Schnakenberg}},
  \bibinfo{journal}{Rev. Mod. Phys.} \textbf{\bibinfo{volume}{48}},
  \bibinfo{pages}{571} (\bibinfo{year}{1976}).

\bibitem[{\citenamefont{White et~al.}(2000)\citenamefont{White, Rubinstein, and
  Kay}}]{white00}
\bibinfo{author}{\bibfnamefont{J.~A.} \bibnamefont{White}},
  \bibinfo{author}{\bibfnamefont{J.~T.} \bibnamefont{Rubinstein}},
  \bibnamefont{and} \bibinfo{author}{\bibfnamefont{A.~R.} \bibnamefont{Kay}},
  \bibinfo{journal}{Trends Neurosci.} \textbf{\bibinfo{volume}{23}},
  \bibinfo{pages}{131} (\bibinfo{year}{2000}).

\bibitem[{\citenamefont{Vilar et~al.}(2002)\citenamefont{Vilar, Kueh, Barkai,
  and Leibler}}]{vilar02}
\bibinfo{author}{\bibfnamefont{J.~M.~G.} \bibnamefont{Vilar}},
  \bibinfo{author}{\bibfnamefont{H.~Y.} \bibnamefont{Kueh}},
  \bibinfo{author}{\bibfnamefont{N.}~\bibnamefont{Barkai}}, \bibnamefont{and}
  \bibinfo{author}{\bibfnamefont{S.}~\bibnamefont{Leibler}},
  \bibinfo{journal}{Proc. Natl. Acad. Sci. USA} \textbf{\bibinfo{volume}{99}},
  \bibinfo{pages}{5988} (\bibinfo{year}{2002}).

\bibitem[{\citenamefont{Paulsson}(2004)}]{paulsson04}
\bibinfo{author}{\bibfnamefont{J.}~\bibnamefont{Paulsson}},
  \bibinfo{journal}{Nature} \textbf{\bibinfo{volume}{427}},
  \bibinfo{pages}{415} (\bibinfo{year}{2004}).

\bibitem[{\citenamefont{Schnakenberg}(1979)}]{schnakenberg79}
\bibinfo{author}{\bibfnamefont{J.}~\bibnamefont{Schnakenberg}},
  \bibinfo{journal}{J. Theor. Biol.} \textbf{\bibinfo{volume}{81}},
  \bibinfo{pages}{389} (\bibinfo{year}{1979}).

\bibitem[{\citenamefont{Goldbeter}(1996)}]{goldbeter}
\bibinfo{author}{\bibfnamefont{A.}~\bibnamefont{Goldbeter}},
  \emph{\bibinfo{title}{Biochemical Oscillations and Cellular Rhythms.}}
  (\bibinfo{publisher}{Cambridge Univ. Press}, \bibinfo{address}{Cambridge},
  \bibinfo{year}{1996}).

\bibitem[{\citenamefont{DeVille et~al.}(in preparation)\citenamefont{DeVille,
  Muratov, and Vanden~Eijnden}}]{dmv}
\bibinfo{author}{\bibfnamefont{R.~E.~L.} \bibnamefont{DeVille}},
  \bibinfo{author}{\bibfnamefont{C.~B.} \bibnamefont{Muratov}},
  \bibnamefont{and}
  \bibinfo{author}{\bibfnamefont{E.}~\bibnamefont{Vanden~Eijnden}}
  (\bibinfo{year}{in preparation}).

\bibitem[{\citenamefont{Shwartz and Weiss}(1995)}]{shwartz}
\bibinfo{author}{\bibfnamefont{A.}~\bibnamefont{Shwartz}} \bibnamefont{and}
  \bibinfo{author}{\bibfnamefont{A.}~\bibnamefont{Weiss}},
  \emph{\bibinfo{title}{Large Deviations for Performance Analysis}}
  (\bibinfo{publisher}{Chapman and Hall}, \bibinfo{address}{London},
  \bibinfo{year}{1995}).

\end{thebibliography}

\end{document}